\documentclass[aps,preprint]{revtex4}%
\usepackage{amsfonts}
\usepackage{amsmath}
\usepackage{amssymb}
\usepackage{graphicx}%
\begin{document}
\title{{\LARGE ENERGY AND ANGULAR MOMENTUM OF DILATON BLACK HOLES}}
\author{Marcelo Samuel Berman$^{1}$}
\affiliation{$^{1}$Instituto Albert Einstein/Latinamerica\ - Av. Candido Hartmann, 575 -
\ \# 17}
\affiliation{80730-440 - Curitiba - PR - Brazil - email: msberman@institutoalberteinstein.org}
\date{(Original: 05 February, 2007. New Version: 04 May, 2008)}

\begin{abstract}
Following a prior paper, we review the results for the energy and angular
momentum of a Kerr-Newman black hole, and then calculate the same properties
for the case of a generalised rotating dilaton of the type derived, without
rotation, by Garfinkle, Horowitz, and Strominger (1991; 1992).\ We show that
there is, as far as it refers only to the energy and angular momentum, an
interaction among the fields, so that, the gravitational and electromagnetic
fields may be obscured by the strength of the scalar field.

(Spanish)

Dando seguimiento a un art\'{\i}culo previo, revisamos los resultados para la
energ\'{\i}a y momento angular de un hoyo negro de Kerr-Newman, y extendemos
el calculo para el caso de un dilaton en rotaci\'{o}n, obtenido a partir del
modelo de Garfinkle, Horowitz, y Strominger (1991; 1992). Mostramos que hay,
en lo que se refiere solamente a la energ\'{\i}a y momento angular, una
interacci\'{o}n entre los campos, de forma que, el gravitacional y el
electromagnetico, puedem ser ocultados por la intensidad del campo escalar.

\end{abstract}
\maketitle

\begin{center}

{\LARGE ENERGY AND ANGULAR MOMENTUM OF DILATON BLACK HOLES}

\bigskip

Marcelo Samuel Berman
\end{center}

\bigskip

{\LARGE I. INTRODUCTION}

\bigskip Astrophysical black holes appear in a wide variety of environments.
We find then, from stellar evolution (X-ray binaries, supernovas, collapsars,
etc), some of intermediate mass or supermassive ones, like those that lie at
the centre of different galaxies. (see, for instance some papers by Noyola and
collaborators, like Noyola et al., 2008; and in books like, Eckart,
Straubmeier and Sch\"{o}del (2005), Lee and Parck (2002), and Kreitler (2006;
2006a; 2006b), and we refer to them for further information).

\bigskip

Astrophysical black holes always have a certain amount of angular momentum
associated to them. It is perhaps this angular momentum which gives rise to
the energetic jets that are observed on many astrophysical systems (see for
instance, Falcone et al. (2008); see also, Pope et al.(2008) on this) such as
X-ray binaries, long gamma-ray bursts and quasars.

\bigskip

For the benefit of readers that are astronomers, we introduce now the concept
of scalar fields and dilatons in a more or less historical perspective. There
are two different situations in which a scalar field has importance: first, as
a time-varying gravitational, and, cosmological, "constants". The first
gravitational theory of that kind was Brans-Dicke (1961). Second, in the
context of string theory. In the latter, the "\textit{dilaton}" is related to
the graviton; in the former, the scalar field is related to the Machian
concept, whereby there is a causally related inertial phenomenon in local
physics, due to the overall distribution of mass in the Universe. In fact,
nowadays it is clear that the gravitational scalar field is very close to the
strings' dilatons. Let us give a trivial example. Wald (1994), suggests that
the problem of black-hole evaporation could be approached by considering a
\textit{dilaton} scalar field in lower dimensional general relativity, with a
"string-inspired" action, 

\bigskip

$S=\frac{1}{2}\pi\int d^{2}x\sqrt{-g}\left[  e^{-2\Phi}\left(  R+4\nabla
a\Phi\nabla^{a}\Phi+4\Lambda^{2}\right)  -\frac{1}{2}\nabla_{a}\phi\nabla
^{a}\phi\right]  $ \ \ \ \ \ \ \ \ \ \ ,

\bigskip

where \ $\Lambda$\ \ , \ \ $\Phi$\ \ \ and \ \ \ $\phi$\ \ \ stand for the
cosmological constant, the \textit{dilaton} field, and a scalar one,
respectively. Classically, the field equations describe the gravitational
collapse of the matter field \ \ $\phi$\ \ \ , \ which yields a black hole. It
could appear to provide testing ideas for the quantum behavior of black holes,
for instance treating \ $\phi$\ \ as a quantum field propagating in a
classical background spacetime, corresponding to the formation of a black hole
by gravitational collapse. \ \ 

\bigskip

Dilaton field-black holes, were studied recently by Vagenas (2003). (See also
Xulu (1998)). Scalar fields, defined by a cosmological constant, plus electric
charge and gravitation, were also the case in a recent paper by Mart\'{\i}nez
and Troncoso(2006), with important cosmological applications (Halpern, 2008).

\bigskip

\bigskip Scalar fields may alter our view of the Universe. Kaluza-Klein
theory, contains a scalar field arising from the pentadimensional 5-5
component of the metric tensor (Wesson, 1999; 2006). Such scalar field,
generally named as \textit{dilatons,} were also identified with inflationary
model's \textit{inflaton} (Collins, Martin and Squires, 1989). String and
brane theories, deal with \textit{dilatons} which play r\^{o}les similar to
the gravitons. String theories have compactified internal space, whose size
arises a \textit{dilaton, }or scalar field. Altogether, it has been claimed
that gravitons interact among themselves and may have also scalar field
companions. Scalar fields disguised under a cosmological "constant" term,
provide clues to dark energy and dark matter models, in addition to the
inflationary ones, plaguing astrophysical and cosmological literature. Scalar
fields may add a little complexity to the vacuum. The four dimensional energy
of the vacuum is a measure of the five dimensional scalar field (Wesson,
2006). For the energy of the vacuum, in connection with gravity and scalar
fields, see, for instance, Berman (2007; 2007a; 2007b; 2008); Faraoni (2004);
Fujii and Maeda (2003).

\bigskip

\bigskip The calculation of energy and angular momentum of black-holes, has,
among others, an important \ astrophysical r\^{o}le, because such objects
remain the ultimate source of energy in the Universe, and \ the amount of
angular momentum is related to the possible amount of extraction of energy
from the b.h.(Levinson, 2006; Kreitler, 2006, 2006a, 2006b). The consequence
for jet production is also of astrophysical interest.

Therefore, Berman(2007), checked whether the calculation of energy and angular
momenta contents for a \ K.N. black hole given by \ Virbhadra, and
Aguirregabiria et al., included the gravitomagnetic contribution. It was seen
that this did not occur. Berman recalculated the energy and angular momenta
formulae, in order that gravitomagnetism enters into the scenario. In the
present text, we advance the theoretical framework, by studying the effect of
a \textit{dilaton} or scalar field, within charged rotating black holes.

\bigskip

{\LARGE II. REVIEW OF PREVIOUS RESULTS}

Chamorro and Virbhadra (1996) have calculated the energy of a spherically
symmetric charged non-rotating \textit{dilaton} black hole, which obeys the metric,

\bigskip

$ds^{2}=A^{-1}dt^{2}-Adr^{2}-Dr^{2}(d\Omega^{2})$ \ \ \ \ \ \ \ \ \ \ \ . \ \ \ \ \ \ \ \ \ \ \ \ \ \ \ \ \ \ \ \ \ \ \ \ \ \ \ \ \ \ \ \ \ (1)

\bigskip

Garfinkle, Horowitz, and Strominger (1991; 1992) departed from a variational
principle which included a scalar field \ $\Phi$\ \ , and the electromagnetic
tensor \ $F_{\alpha\beta}$\ \ , in addition to the Ricci scalar \ \ $R$\ \ ,
to wit,

\bigskip

$\delta\int\left[  -R+2\left(  \nabla\Phi\right)  ^{2}+e^{-2\beta\Phi}%
F^{2}\right]  \sqrt{-g}d^{4}x=0$ \ \ \ \ \ \ . \ \ \ \ \ \ \ \ \ \ \ \ \ \ \ \ \ \ \ \ \ \ \ \ \ \ \ (2)

\bigskip

The resultant field equations are:

\bigskip

$\nabla_{j}\left(  e^{-2\beta\Phi}F^{jk}\right)  =0$ \ \ \ \ \ \ \ \ \ \ \ \ \ \ \ \ \ \ ,

\bigskip$\nabla^{2}\Phi+\frac{\beta}{2}e^{-2\beta\Phi}F^{2}=0$ \ \ \ \ \ \ \ \ \ \ \ ,

$R_{ij}=2\nabla_{i}\Phi\nabla_{j}\Phi+2e^{-2\beta\Phi}F_{ia}F_{j}^{a}-\frac
{1}{2}g_{ij}e^{-2\beta\Phi}F^{2}$ \ \ \ \ \ \ \ \ ,

\bigskip and, the \textit{dilaton} was described by the following solution:

\bigskip

$e^{-2\Phi}=\left[  1-\frac{r_{-}}{r}\right]  ^{\frac{1-\sigma}{\beta}}$
\ \ \ \ \ \ \ \ \ \ . \ \ \ \ \ \ \ \ \ \ \ \ \ \ \ \ \ \ \ \ \ \ \ \ \ \ \ \ \ \ \ \ \ \ \ \ \ \ \ \ \ \ \ \ \ \ \ \ \ \ (3)

\bigskip

The sign of \ $\beta$\ \ only influences the sign of \ $\Phi$\ ; we are going
therefore to take \ $\beta>0$\ \ .

\bigskip

The usual Coulomb interaction is given by,

\bigskip

$F_{0r}=\frac{Q}{r^{2}}$ \ \ \ \ \ \ \ \ \ \ \ . \ \ \ \ \ \ \ \ \ \ \ \ \ \ \ \ \ \ \ \ \ \ \ \ \ \ \ \ \ \ \ \ \ \ \ \ \ \ \ \ \ \ \ \ \ \ \ \ \ \ \ \ \ \ \ \ \ \ \ \ \ \ \ \ \ \ \ \ \ \ \ \ \ \ \ \ \ \ \ \ \ \ \ \ \ (4)\ 

\bigskip

The metric coefficients are,

\bigskip

$A^{-1}=\left(  1-\frac{r_{+}}{r}\right)  \left(  1-\frac{r_{-}}{r}\right)
^{\sigma}$ \ \ \ \ \ \ \ \ \ \ , \ \ \ \ \ \ \ \ \ \ \ \ \ \ \ \ \ \ \ \ \ \ \ \ \ \ \ \ \ \ \ \ \ \ \ \ \ \ \ \ \ \ \ \ \ \ \ \ \ \ \ \ \ \ \ \ \ \ \ \ \ (5)

\bigskip

and,

\bigskip

$D=\left(  1-\frac{r_{-}}{r}\right)  ^{1-\sigma}$ \ \ \ \ \ \ \ \ \ \ . \ \ \ \ \ \ \ \ \ \ \ \ \ \ \ \ \ \ \ \ \ \ \ \ \ \ \ \ \ \ \ \ \ \ \ \ \ \ \ \ \ \ \ \ \ \ \ \ \ \ \ \ \ \ \ \ \ \ \ \ \ \ \ \ \ \ \ \ \ \ \ \ \ \ \ (6)

\bigskip

In the above, we have made use of the following constraints and/or definitions,

\bigskip

$\sigma=\frac{1-\beta^{2}}{1+\beta^{2}}$ \ \ \ \ \ \ \ \ \ . \ \ \ \ \ \ \ \ \ \ \ \ \ \ \ \ \ \ \ \ \ \ \ \ \ \ \ \ \ \ \ \ \ \ \ \ \ \ \ \ \ \ \ \ \ \ \ \ \ \ \ \ \ \ \ \ \ \ \ \ \ \ \ \ \ \ \ \ \ \ \ \ \ \ \ \ \ \ \ \ \ \ \ \ \ \ \ (7)

\bigskip

$r_{+}+\sigma r_{-}\equiv2M$ \ \ \ \ \ \ \ \ \ \ \ \ \ \ . \ \ \ \ \ \ \ \ \ \ \ \ \ \ \ \ \ \ \ \ \ \ \ \ \ \ \ \ \ \ \ \ \ \ \ \ \ \ \ \ \ \ \ \ \ \ \ \ \ \ \ \ \ \ \ \ \ \ \ \ \ \ \ \ \ \ \ \ \ \ \ \ \ \ (8)

\bigskip

$r_{+}r_{-}\equiv Q^{2}\left(  1+\beta^{2}\right)  $
\ \ \ \ \ \ \ \ \ \ \ \ \ \ . \ \ \ \ \ \ \ \ \ \ \ \ \ \ \ \ \ \ \ \ \ \ \ \ \ \ \ \ \ \ \ \ \ \ \ \ \ \ \ \ \ \ \ \ \ \ \ \ \ \ \ \ \ \ \ \ \ \ \ \ \ \ \ \ \ \ \ \ (9)

\bigskip

It can be seen that \ \ $\beta$\ \ \ rules the relative intensity among the
three fields, gravitational, electromagnetic and scalar.

\bigskip

In determining the mass and angular momenta of a given asymptotically flat
space-time, there are in the literature a number of specific procedures, such
as the ADM-Mass, or related pseudo-tensor and gravitational superpotential
theories. Some complexes (Landau-Lifshitz, Papapetrou, Weinberg, etc) are
usually employed. An important step towards the freedom on their use, has been
the calculation of Aguirregabiria, Chamorro and Virbhadra (1996), showing that
most of them yield the same result when applied to a large class of metrics.

\bigskip

The lesson given by Berman(2007), was that when energy or momentum were
calculated, it sufficed to take the charge contribution, leaving \ \ $M=0$\ ,
and, at the end of pseudotensorial calculation, \ making the following transformation:

\bigskip

$Q^{2}\rightarrow\left[  Q^{2}+M^{2}+P^{2}\right]  $\ \ \ \ \ \ \ \ \ \ , \ \ \ \ \ \ \ \ \ \ \ \ \ \ \ \ \ \ \ \ \ \ \ \ \ \ \ \ \ \ \ \ \ \ \ \ \ \ \ \ \ \ \ \ \ \ \ \ \ \ \ \ \ \ \ \ \ \ \ \ \ \ \ \ \ (10)

\bigskip

where \ \ \ $P$\ \ \ \ stands for the magnetic charge\ (if there is some).\ 

\bigskip

\bigskip Of course, there should be made room for the inertial content,
\ \ $Mc^{2}$\ \ in the case of the energy, and \ $aM$\ \ , in the case of
rotating black hole's angular momentum: these two terms were the total energy
or momentum, when \ $r\rightarrow\infty$\ \ \ .

\bigskip

For instance, if the electric energy of Reissner-Nordstr\"{o}m's black hole
was given by \ $\ -\frac{Q^{2}}{2r}$\ \ , the total contributions for the
energy content would be written as,

\bigskip

$E_{RN}=Mc^{2}-\frac{\left[  Q^{2}+M^{2}+P^{2}\right]  }{2r}$%
\ \ \ \ \ \ \ \ \ . \ \ \ \ \ \ \ \ \ \ \ \ \ \ \ \ \ \ \ \ \ \ \ \ \ \ \ \ \ \ \ \ \ \ \ \ \ \ \ \ \ \ \ \ \ \ \ \ \ \ \ \ \ \ \ \ \ \ \ \ \ \ \ \ \ \ (11)

\bigskip

When a scalar field of the above form enters into the scene, Chamorro and
Virbhadra (1996) found, by pseudo-tensor calculations, for the electric
contribution, the term, $-\frac{Q^{2}}{2r}\left(  1-\beta^{2}\right)  $\ .
Therefore, by means of our rule, we have the complete formula as given by,

\bigskip

$E=$\ $Mc^{2}-\frac{\left[  Q^{2}+M^{2}+P^{2}\right]  }{2r}\left(  1-\beta
^{2}\right)  $\ \ \ \ \ \ \ \ \ . \ \ \ \ \ \ \ \ \ \ \ \ \ \ \ \ \ \ \ \ \ \ \ \ \ \ \ \ \ \ \ \ \ \ \ \ \ \ \ \ \ \ \ \ \ \ \ \ \ \ \ \ \ \ \ \ \ (12)

\bigskip

We now turn our attention to the rotating charged situation. By analogy with
the above case, consider that, for a K.N. black hole, the metric may be given
in Cartesian coordinates by:

\ 

\bigskip\hfill$ds^{2}=dt^{2}-dx^{2}-dy^{2}-dz^{2}-\frac{2\left[  M-\frac
{Q^{2}}{2r_{0}}\right]  r_{0}^{3}}{r_{0}^{4}+a^{2}z^{2}}\cdot\bar{F}^{2}$
,\hfill(13)

\bigskip\noindent while,

\bigskip\hfill$\bar{F}=dt+\frac{Z}{r_{0}}dz+\frac{r_{0}}{\left(  r_{0}%
^{2}+a^{2}\right)  }\left(  xdx+ydy\right)  +\frac{a\left(  xdy-ydx\right)
}{a^{2}+r_{0}^{2}}$,\hfill(14)

\bigskip\hfill$r_{0}^{4}-\left(  r^{2}-a^{2}\right)  r_{0}^{2}-a^{2}z^{2}%
=0$,\hfill(15)

\bigskip\noindent and,

\bigskip\hfill$r^{2}\equiv x^{2}+y^{2}+z^{2}$ .\hfill(16)

\bigskip

In the above, \ $M$\ , \ $Q$\ \ \ and \ "$a$" \ stand respectively for the
mass, electric charge, and the rotational parameter, which has been shown to
be given by:

\bigskip\hfill$a=\frac{J_{TOT}}{M}$,\hfill(17)

\bigskip\noindent where \ $J_{TOT}$\ \ stands for the total angular momentum
of the system, in the limit \ $R\rightarrow\infty$\ . As Berman (2007)
described in his recent paper, we may keep the electric energy calculations by
Virbhadra (1990; 1990a; 1990b) and Aguirregabiria et al. (1996), and, by
applying the transformation (10), obtaining, for the energy and angular
momenta, the formulae of \ Berman(2007):

\bigskip

\bigskip\hfill$\left(  P_{0}\right)  _{KN}=M-\left[  \frac{Q^{2}+M^{2}+P^{2}%
}{4\varrho}\right]  \left[  1+\frac{\left(  a^{2}+\varrho^{2}\right)
}{a\varrho}arctgh\left(  \frac{a}{\varrho}\right)  \right]  $
\ \ \ \ \ \ ,\hfill(18)

\bigskip\hfill$P_{1}=P_{2}=P_{3}=0$ \ \ \ \ \ \ \ \ \ \ \ \ \ \ \ .\hfill(19)

\bigskip Likewise, if we apply:

$J^{(3)}=\int\left[  x^{1}p_{2}-x^{2}p_{1}\right]  d^{3}x$ \ \ \ \ \ \ \ \ ,

\bigskip

we find,\hfill

\bigskip

\bigskip$\left(  J^{(3)}\right)  _{KN}=aM-\left[  \frac{Q^{2}\text{ }+\text{
}M^{2}+P^{2}}{4\varrho}\right]  a\left[  1-\frac{\rho^{2}}{a^{2}}%
+\frac{\left(  a^{2}+\varrho^{2}\right)  ^{2}}{a^{3}\varrho}arctgh\left(
\frac{a}{\varrho}\right)  \right]  $ \ \ \ \ \ . \ \ \ \ \ \ \ \ \ \ \ \ \ \ \ \ \ (20)

\bigskip

$J^{(1)}=J^{(2)}=0$ \ \ \ \ \ \ \ \ \ \ \ \ \ \ .\hfill(21)

\bigskip

{\LARGE III. ROTATING KERR-NEWMAN DILATON ENERGY-MOMENTUM}

\bigskip

\bigskip We now are able to write the corresponding result, for a
\textit{dilaton} \ Kerr-Newman black hole's energy and momenta, \bigskip
\noindent where, \ the linear momentum densities are given by:\ 

\bigskip$p_{1}=-2\left(  1-\beta^{2}\right)  \left[  \frac{\left(  Q^{2}%
+M^{2}+P^{2}\right)  \rho^{4}}{8\pi(\rho^{4}+a^{2}z^{2})^{3}}\right]
ay\rho^{2}$ ,\hfill

\bigskip$p_{2}=-2\left(  1-\beta^{2}\right)  \left[  \frac{\left(  Q^{2}%
+M^{2}+P^{2}\right)  \rho^{4}}{8\pi(\rho^{4}+a^{2}z^{2})^{3}}\right]
ax\rho^{2}$ ,\hfill

\bigskip$p_{3}=0$ ,\hfill

\bigskip\noindent while the energy density is given by:

\bigskip$\mu=\left(  1-\beta^{2}\right)  \left[  \frac{\left(  Q^{2}%
+M^{2}+P^{2}\right)  \rho^{4}}{8\pi(\rho^{4}+a^{2}z^{2})^{3}}\right]  \left(
\rho^{4}+2a^{2}\rho^{2}-a^{2}z^{2}\right)  $ \ \ \ \ .\hfill

\bigskip The energy and angular momenta are then,

\bigskip\bigskip$\left(  P_{0}\right)  _{dilaton}=M-\left[  \frac{Q^{2}%
+M^{2}+P^{2}}{4\varrho}\right]  \left[  1+\frac{\left(  a^{2}+\varrho
^{2}\right)  }{a\varrho}arctgh\left(  \frac{a}{\varrho}\right)  \right]
\left(  1-\beta^{2}\right)  $ \ \ \ , \ \ \ \ \ \ \ \ \ \ \ \ \ \ \ \ \ \ \ (22)

\bigskip$P_{1}=P_{2}=P_{3}=0$ \ \ \ \ \ \ \ \ \ \ \ \ \ \ \ .\hfill(23)

$\left(  J^{(3)}\right)  _{dilaton}=aM-\left[  \frac{Q^{2}\text{ }+\text{
}M^{2}+P^{2}}{4\varrho}\right]  a\left[  1-\frac{\rho^{2}}{a^{2}}%
+\frac{\left(  a^{2}+\varrho^{2}\right)  ^{2}}{a^{3}\varrho}arctgh\left(
\frac{a}{\varrho}\right)  \right]  \left(  1-\beta^{2}\right)  $ \ \ ,\ \ \ \ \ \ (24)

\bigskip

and,

\bigskip

$J^{(1)}=J^{(2)}=0$ \ \ \ \ \ \ \ \ \ \ \ \ \ \ .\hfill(25)

\bigskip

\bigskip In the above, we define \ $\rho$\ \ as the positive root of equation ,

\bigskip

\bigskip$\frac{x^{2}+y^{2}}{\varrho^{2}+a^{2}}+\frac{z^{2}}{\varrho^{2}}=1$
\ \ \ \ \ \ \ \ \ \ \ \ \ \ .\hfill\hfill(26)

\bigskip Relations (23) and (25), "validate"\ the coordinate system chosen for
the present calculation: it is tantamount to the choice of a center-of-mass
coordinate system in Newtonian Physics, or the use of comoving observers in Cosmology.

We made the following transformation,

\bigskip

$Q^{2}\rightarrow\left[  Q^{2}+M^{2}+P^{2}\right]  \left(  1-\beta^{2}\right)
$ \ \ \ \ \ \ \ \ \ \ \ \ \ , \ \ \ \ \ \ \ \ \ \ \ \ \ \ \ \ \ \ \ \ \ \ \ \ \ \ \ \ \ \ \ \ \ \ \ \ \ \ \ \ \ \ \ \ \ \ \ \ \ \ \ \ \ (26b)

\bigskip

which takes us from the black hole charge contribution, to the total scalar
field -- electromagnetic charges -- gravitation field contributions, which
constitute the \textit{dilaton}\ Kerr-Newman black hole!!!\ 

\bigskip That our method "works", is a question of applying, say, some
superpotential calculations. In Berman (2007), we have supported this method
for the case \ $\beta=0$\ \ . There is no reason not to generalise it to
\ \ $\beta\neq0$\ \ \ cases. But, again, we can be sure that our formulae
keeps intact the following physical good properties:

\bigskip

1) gravitomagnetic effects are explicit;

2) the triple interaction, among scalar, gravitational and electromagnetic
fields becomes evident, as far as energy and angular momenta are concerned; and

3) when \ $\beta=1$\ \ , the scalar field neutralizes the other interactions;
if \ $\beta<1$\ \ , the neutralization is only partial.\ 

\bigskip

\bigskip By considering an expansion of the arcth($\frac{a}{\varrho}$)
function, in terms of increasing powers of the parameter \textquotedblright%
$a$\textquotedblright, and by neglecting terms $\left(  \frac{a}{\rho}\right)
^{3+n}\simeq0$, ( \ with \ $n=0$, \ $1$\ \ ,\ $2$ , . . . )\ \ we find the
energy of a slowly rotating \textit{dilaton }Kerr-Newman black-hole,

\bigskip

\bigskip$E\simeq M-\left[  \frac{Q^{2}+M^{2}+P^{2}}{R}\right]  \left[
\frac{a^{2}}{3R^{2}}+\frac{1}{2}\right]  \left(  1-\beta^{2}\right)  $
\ \ \ \ \ \ \ \ \ ,\hfill(27)

\bigskip

\noindent where \ $\varrho\rightarrow R$ \ , if $\ \ \ a\rightarrow$\ $0$
\ \ , \ according to (26).

We can interpret the terms $\frac{\left(  Q^{2}+P^{2}\right)  a^{2}\left(
1-\beta^{2}\right)  }{3R^{3}}$ \ and $\frac{M^{2}a^{2}\left(  1-\beta
^{2}\right)  }{3R^{3}}$ \ as the magnetic and gravitomagnetic energies caused
by rotation.

\bigskip

\bigskip\bigskip Expanding the \ $arctgh$ \ function in powers of ($\frac
{a}{\varrho}$)\ , and retaining \ up to third power, we find the slow rotation
angular momentum:

\bigskip$J^{(3)}\cong aM-2\left[  Q^{2}+M^{2}+P^{2}\right]  a\left[
\frac{a^{2}}{5R^{3}}+\frac{1}{3R}\right]  \left(  1-\beta^{2}\right)  $
.\hfill(28)

\bigskip

In the same approximation, we would find:

\bigskip$\mu\cong\left[  \frac{Q^{2}+M^{2}+P^{2}}{4\pi R^{4}}\right]  \left[
\frac{a^{2}}{R^{2}}+\frac{1}{2}\right]  \left(  1-\beta^{2}\right)  $ .\hfill(29)

\bigskip

The above formula could be also found by applying directly the definition, \ \ \ \ 

\bigskip$\mu=\frac{dP_{0}}{dV}=\frac{1}{4\pi R^{2}}\frac{dP_{0}}{dR}$ .\hfill(30)

\bigskip

\bigskip We further conclude that we may identify the gravitomagnetic
contribution to the energy and angular momentum of the \textit{dilaton }K.N.
black hole, for the slow rotating case, as:

\bigskip$\Delta E\cong-\frac{M^{2}a^{2}}{3R^{3}}\left(  1-\beta^{2}\right)  $
,\hfill(31)

\bigskip\noindent and,

\bigskip$\Delta J\cong-2M^{2}\left[  \frac{a^{3}}{5R^{3}}+\frac{a}{3R}\right]
\left(  1-\beta^{2}\right)  \approx-\frac{2M^{2}a}{3R}\left(  1-\beta
^{2}\right)  $ ,\hfill(32)

\bigskip\noindent as can be easily checked by the reader.

\bigskip

{\LARGE IV. THE METRIC ELEMENT FOR THE K-N DILATON}

\bigskip

We may succeed in obtaining the correct metric for the K-N \textit{dilaton}
black hole, by requiring that:

\bigskip

\begin{quote}
\textit{1. when \ \ }$\beta=0$\textit{\ \ , we must retrieve back K-N original
metric (13);}

\textit{2. when \ \ }$a=0$\textit{\ \ , we should obtain Garfinkle et al's
metric (1);}

\textit{3. when \ \ }$\beta=a=0$\textit{\ \ , we should reduce to
Reissner-Nordstr\"{o}m's metric;}

\textit{4. in the reversed order, we must keep the same relationship, either
between \ Reissner-Nordstr\"{o}m's and K-N metric's, or, between Garfinkle et
al's and our new metric, to be presented below.}

\textit{5. Chamorro and Virbhadra's result, should be derived from our new
result, provided that we take care of transformation (26b).}

\textit{6. the metric to be found, should be the simplest one to obey the
above requirements.}
\end{quote}

{\tiny \bigskip}

\bigskip We now present the metric:

\bigskip

$ds^{2}=Rdt^{2}-Sdr^{2}-Dr^{2}d\Omega^{2}-\frac{2\left[  M\sqrt{1-\beta^{2}%
}-\frac{\left[  Q^{2}\right]  \left(  1-\beta^{2}\right)  }{2r_{0}}\right]
r_{0}^{3}}{r_{0}^{4}+a^{2}z^{2}}\cdot\bar{F}^{2}$
\ \ \ \ \ \ \ \ \ \ \ \ \ ,\hfill(33)

\bigskip

where, \ \ $\bar{F}$ \ \ is given by relations (14), (15) and (16), and, \ 

\bigskip

$R\equiv A^{-1}+\Gamma$ \ \ \ \ \ \ \ \ \ \ \ \ \ \ \ \ \ \ \ \ , \ \ \ \ \ \ \ \ \ \ \ \ \ \ \ \ \ \ \ \ \ \ \ \ \ \ \ \ \ \ \ \ \ \ \ \ \ \ \ \ \ \ \ \ \ \ \ \ \ \ \ \ \ \ \ \ \ \ \ \ \ \ \ \ \ \ \ \ \ \ \ \ \ \ \ \ \ (34)

\bigskip

$S\equiv A+\frac{\Gamma}{\Gamma-1}$\ \ \ \ \ \ \ \ \ \ \ \ \ \ \ \ \ \ \ \ , \ \ \ \ \ \ \ \ \ \ \ \ \ \ \ \ \ \ \ \ \ \ \ \ \ \ \ \ \ \ \ \ \ \ \ \ \ \ \ \ \ \ \ \ \ \ \ \ \ \ \ \ \ \ \ \ \ \ \ \ \ \ \ \ \ \ \ \ \ \ \ \ \ \ \ \ \ \ (35)

\bigskip

$\Gamma\equiv\frac{2M}{r}\sqrt{1-\beta^{2}}-\frac{\left[  Q^{2}\right]
\left(  1-\beta^{2}\right)  }{r^{2}}$ \ \ \ \ \ \ \ \ \ \ \ \ \ \ \ . \ \ \ \ \ \ \ \ \ \ \ \ \ \ \ \ \ \ \ \ \ \ \ \ \ \ \ \ \ \ \ \ \ \ \ \ \ \ \ \ \ \ \ \ \ \ \ \ \ \ \ \ \ \ \ \ \ \ (36)

\bigskip

The above metric represents our \textit{dilaton}.\bigskip

{\LARGE V. CONCLUDING REMARKS}

It is important to notice that the contributed energy, due to the scalar field
is given by the term \ \ \ $\frac{\beta^{2}}{2r}\left[  M^{2}+Q^{2}%
+P^{2}\right]  >0$\ \ \ , \ but the corresponding energy density is negative,
given by \ $-\left[  \frac{\beta^{2}\left[  M^{2}+Q^{2}+P^{2}\right]  }{8\pi
R^{4}}\right]  $\ \ . \ This negative energy density, is the trademark of the
scalar field. It must be remarked that all of our results do not match with
Chamorro and Virbhadra's, except in the particular case when \ \ $M=P=0$\ . Of
course, those authors only examined the Reissner-Nordstr\"{o}m's
\ \textit{dilaton}\ , a non-rotating black hole. We also found that there is a
relative interaction between matter, charges and the scalar field.

\bigskip

We remember that the terms \ \ $Mc^{2}$\ \ and \ \ $aM$\ \ \ which appear
respectively, in the energy and momentum formulae, refer to inertia and not to
gravitation: thus, they refer to Special Relativity. We have found also, that
the scalar field reduces the self-energies, of gravitation and electromagnetic
origin, by a factor \ $\left(  1-\beta^{2}\right)  $\ . This fact remains an
important feature of the present derivations, since we may think of a kind of
new Equivalence\ Principle \ under the possibility that not only acceleration
is equivalent to a gravitational field, but the kind of neutralization we have
studied points to a way of eliminating gravity at small scales, at least, by
means of a scalar field.\ We have been accused of a very "lousy" use of the
neutralization property cited above; however, we must take care, because we
are only dealing with the energy momentum concept, and the Physics\ of the
problem has a lot more to say, in addition to energy considerations. For
instance, the \ $\beta$\ \ parameter has been making a shift in the location
of the horizons but may not be the only ruler of the intensity among the three
fields, when we are dealing with other physically related properties.

\bigskip

{\LARGE Acknowledgements}

\bigskip

\bigskip Two anonymous referees have given invaluable advices that have made a
difference, in the last version. I am grateful to them, and also to a private
communication by K.S. Virbhadra, stating that in the paper by Aguirregabiria
et al. (1996), the correct function is $arctgh$ and not \ $arctan$ \ \ , as
stated erroneously there. (See relations (18), (20), (22), (24) of the present
paper). The author also thanks his intellectual mentors, Fernando de Mello
Gomide and M. M. Som, and also to Marcelo Fermann Guimar\~{a}es, Nelson Suga,
Mauro Tonasse, Antonio F. da F. Teixeira, and for the encouragement by Albert,
Paula and Geni.

\bigskip

\bigskip{\Large References}

Aguirregabiria, J.M.; Chamorro, A.; Virbhadra, K.S.(1996) - \textit{Gen. Rel.
and Grav. }\textbf{28}, 1393.

Berman, M.S. (2007) - \textit{Gravitomagnetism and Angular Momenta of
Black-Holes} - Revista Mexicana the Astronomia y Astrof\'{\i}sica,
\textbf{43}, 2. For a preliminary report, see: \textit{Los Alamos Archives} http://arxiv.org/abs/physics/0608053

Berman, M.S. (2007a) - \textit{Introduction to General Relativity and the
Cosmological Constant Problem} - Nova Science ,N.Y.

Berman, M.S. (2007b) - \textit{Introduction to General Relativistic and
Scalar-Tensor Cosmologies} - Nova Science , N.Y.

Berman, M.S. (2008) - \textit{A Primer in Black Holes, Mach%
\'{}%
s Principle, and Gravitational Energy} - Nova Science, New York.

\bigskip\ Brans, C.; Dicke, R.H. (1961) - Physical Review, \textbf{124}, 925.

Chamorro, A.; Virbhadra, K.S. (1996) - International Journal \ Modern Physics,
\textbf{D5}, 251.

\bigskip Collins, P.D.B.; Martin, A.D.; Squires, E.J. (1989) -
\textit{Particle Physics and Cosmology}, Wiley, New York.

Eckart, A.; Straubmeier, C.; Sch\"{o}del,R. (2005) - \textit{The Black Hole at
the Center of the Milky Way. }World Scientific, Singapore.

Falcone, A. D. ; Williams, D. A.; Baring, M. G.; Blandford, R.;
Connaughton,V.; Coppi, P.; Dermer, C.; Dingus, B.; Fryer, C.; Gehrels, N.;
Granot, J.; Horan, D.; Katz, J. I.; Kuehn, K.; Meszaros, P.; Norris, J.; Saz
Parkinson, P.; Peer, A.; Ramirez-Ruiz, E.; Razzaque, S.; Wang, X.; Zhang. B.
(2008) - \textit{The Gamma Ray Burst section of the White Paper on the Status
and Future of Very High Energy Gamma Ray Astronomy: A Brief Preliminary Report
. }http://arxiv.org/abs/0804.2256 \ .

Faraoni, V. (2004) - \textit{Cosmology in Scalar Tensor Gravity, }Kluwer
Academic Publishers, Dordrecht.

Fujii, Y.; Maeda, K-I. (2003) - \textit{The Scalar Tensor Theory of
Gravitation, }CUP, Cambridge.

\begin{description}
\item Garfinkle, D.; Horowitz, G.T.; Strominger, A. (1991) - Physical Review,
\textbf{D43}, 3140.
\end{description}

\bigskip Garfinkle, D.; Horowitz, G.T.; Strominger, A. (1992) - Physical
Review, \textbf{D45}, 3888.

Halpern, P. (2008) - \textit{Energy Distribution of a Charged Black Hole with
a Minimally Coupled Scalar Field - }Pre-print - \textit{Los Alamos Archives}
http://arxiv.org/abs/gr-qc/0705.0720 .

Kreitler, P. (2006) - \textit{Trends in Black-Hole Research}, Nova Science,
New York.

Kreitler, P. (2006a) - \textit{New Developments in Black-Hole Research}, Nova
Science, New York.

Kreitler, P. (2006b) - \textit{Focus in Black-Hole Research}, Nova Science,
New York.

Lee, H.K.; Parck, M.-G. (2002) - \textit{Black Hole Astrophysics 2002.}
(Proceedings of the 6$^{th}$ APCTP Winter School), World Scientific, Singapore.

Levinson, A. (2006) - Chapter 4, in \textit{Trends in Black Hole Research},
edited by P. V. Kreitler, Nova Science, New York.

Mart\'{\i}nez, C.; Troncoso, R. (2006) - Physical Review, \textbf{D74}, 064007.

Noyola, E.; Gebhardt K.; Bergmann, M. (2008) - \textit{Gemini and Hubble Space
Telescope Evidence for an Intermediate Mass Black Hole in omega Centauri},
http://arxiv.org/abs/0801.2782 .

Pope, E.; Pittard, J.; Hartquist, T.; Falle, S. (2008) -  \textit{The
generation of optical emission-line filaments in galaxy clusters ,
}http://arxiv.org/abs/0801.2164 .

Vagenas, E.C. (2003) - International Journal of Modern Physics, \textbf{A18, }5949.

\begin{description}
\item Virbhadra, K.S. (1990) - \textit{Phys. Rev.} \textbf{D41}, 1086.

\item Virbhadra, K.S. (1990a) - \textit{Phys. Rev.} \textbf{D42}, 2919.

\item Virbhadra, K.S. (1990b) - \textit{Phys. Rev.} \textbf{D42}, 1066.

\item Xulu, S.S. (1998) - International Journal of Modern Physics,
\textbf{D7}, 773.

\item Wald, R.M. (1994) - \textit{Quantum Field Theory in Curved Spacetime and
Black Hole Thermodynamics, }The University of Chicago Press, Chicago.

\item Wesson, P.S. (1999) - \textit{Space-Time-Matter (Modern Kaluza, Klein
Theory)} , World Scientific, Singapore.

\item Wesson, P.S. (2006) - \textit{Five Dimensional Physics}, World
Scientific, Singapore.
\end{description}

\end{document}